# Analyzing CDR/IPDR data to find People Network from Encrypted Messaging Services

**Adya V. Joshi**
Howard High School
Ellicott City, MD, USA
adyajoshi@gmail.com

**Madan Oberoi**
Special Commissior of Police
Special Cell and Technology Cell
Delhi Police, Delhi, India

**Ranjan Bose**
Indian Institute of Technology Delhi
New Delhi, India

*Abstract*—Criminals are increasingly using mobile based communication applications, like WhatsApp, that have end-to-end encryption to connect to each other. This makes traditional analysis of call graphs, or traffic analysis, virtually impossible and so is a hindrance for law enforcement personnel. Old methods of traffic analysis have been rendered useless and criminals, including arms dealers and terrorists, are able to engage in criminal activity undetected by police. At present, law enforcement must use extensive manual effort to parse data provided by cell companies to extract information. We have built a system that analyses cellular GPRS metadata and builds a profile and finds potential call connections explicitly which are implicit in the dataset. This paper describes our approach and system in detail and includes results of our evaluation using an anonymized dataset from Delhi Police. Our system permits call graph analysis to be done, and significantly reduces the time needed from the data analysis process.

I. INTRODUCTION

In the past few years, people all over the world have rapidly adopted Internet based messaging applications, like WhatsApp [5], Facebook Messenger[4], Rakuten Viber [5], WeChat [7] etc., because of the cost savings they provide. Today over a billion people are using these messaging service to connect within and outside their social network. According to Statista [11], WhatsApp [5] is the most popular mobile messaging app in the world. Facebook Messenger [4] ranked second, followed by Chinese services QQ Mobile and WeChat [7]. Other popular messaging apps include Skype, Viber, LINE, BBM, Telegram and Kakaotalk. Facebook owns the top 2 mobile messaging apps – WhatsApp and Facebook Messenger.

However, it is being increasingly observed that criminals are also adopting these applications since they often provide end-to-end encryption that makes it difficult for law enforcement agencies to intercept them [1]. Traditional methods of mobile traffic analysis have been rendered useless due to the encryption and criminals are able to engage in criminal activity undetected by police. Specifically, given a potential criminal suspect, the law enforcement agencies have no way of identifying their network by analyzing the call graph generated from their mobile phones, since no phone calls are being made. WhatsApp, or other encrypted messaging services, are used instead.

Governments in various countries have expressed concerns that encrypted messaging services can be a threat to national security as it can be used by terrorists and other criminals to harm the society [2][10]. Some countries are even changing laws requiring messaging service providers to provide law enforcement agencies with access to encrypted messages [8][9]. Hence, there is an urgent need to develop new techniques to analyze mobile phone logs to be able to connect potential actors in criminal activities who might be using encrypted messaging services for their criminal activities.

Call Detail Records (CDR) and Internet Protocol Detail Records (IPDR) help track details of a telecommunication call or message generated by a phone device. These logs contain metadata that describe details of a specific call, like calling phone number, destination port, start date/time, end date/time etc. Law enforcement in Delhi has been analyzing CDR and IPDR logs to track potential criminals. This includes GPRS CDRs. While this specifically refers to 2G data related calls, we use the term loosely to describe all CDRs that cover "data" calls that provide the IP connectivity to the handset. However, the current process requires lot of manual effort and is a long and tedious process. Police must go through folders of data provided by cell companies to find one file of metadata. After this they must go through many lines of data, and extract information by hand. This information is typically limited to strings that indicate what website a suspect visited, or strings that might have been part of files locally opened etc. It does not include any attempt at constructing call graphs.

We have developed a system that analyzes mobile Call Data Record (CDR) GPRS and IPDR datasets to generate a profile of the mobile user. It also correlates the mobile users with other users of the encrypted messaging service to make explicit the implicit call graphs in this data. Our key insight is the following. When two suspects are talking to each other on WhatsApp, analyzing their GPRS CDRs will reveal that both are connected to WhatsApp. The reverse is of course not true. Just because two people are connected to WhatsApp at the same time does not mean that they are talking to one another. However, the more two people are on WhatsApp at the same time, the higher the probability that they are talking to one another. This means that by correlating two GPRS CDRs, we can see how often two suspects are using the same service at the same time, and potentially inferring that they are connected to each other.

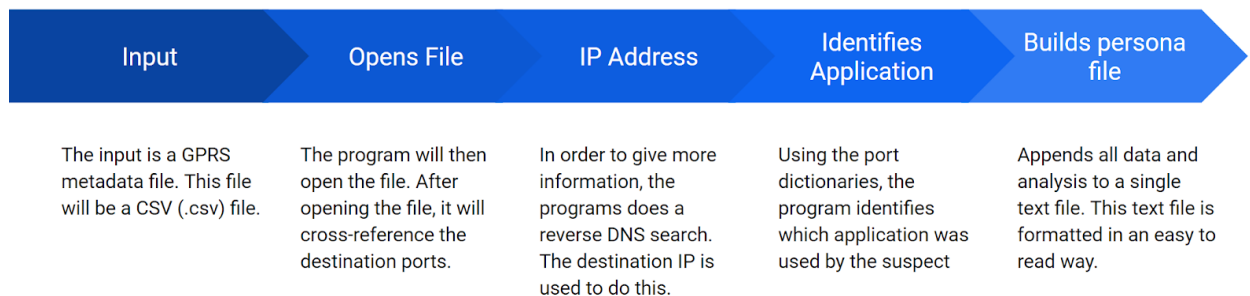

Figure 1: Persona Generation Process

This mobile dataset was obtained by anonymizing data of criminals collected by the special cell unit of Delhi Police [3]. No personally identifiable information (PII) data was included in the logs that were studied for this project. Hence, no prior information of the criminal's profile was available for our system design. Our system significantly reduces the time from the data gathering and analysis process.

In this paper, we describe our approach and the system in detail. Section II describes the related work. Section III describes the system design approach and the results of our evaluation. Section IV describes planned future work that will better this software..

## II. RELATED WORK

### A. CDR Logs Analysis

Teng and Chou [12] have proposed a graphical analytics approach to mining CDR logs to determine potential communities. Zhou et. al. [13] have analyzed CDRs to deduce social attribute from the calling behavior. However, this approach will not work on encrypted messaging services as the connection between the two users cannot be determined from the call logs. The only information the logs provide us is the port number of the application to which the mobile user connected.

### B. WhatsApp Messaging Application

Anglano [17] in his analysis used many types of artifacts generated on devices by WhatsApp in correlation to find pieces of information like contact lists, chronology of messages, deleted contacts, and deleted messages. This information could not be gathered by analyzing these artifacts on their own. This research was limited to Android devices.

Researchers have also reviewed encrypted messaging applications, like WhatsApp to study their impact on human behavior. Montag et. al. [15] conducted demographic analysis of WhatsApp usage. They observed that WhatsApp is about 20% of a typical daily smartphone usage and females, younger people, and those with extraversion tend to use it more than others. The study was conducted by collecting data directly through the subjects' smartphone and analyzing said data. Church et. al. [16] have compared benefits of WhatsApp with traditional SMS technology for mobile communication. These studies have however been limited to mobile data and not CDR logs.

### C. Mobile Data Analysis for Networks

Researchers have analyzed and correlated mobile data logs to determine social networks. Eagle et. al. [14] demonstrated that friendships can be predicted based on the observational mobile data alone, where friend dyads demonstrate distinctive temporal and spatial patterns in their physical proximity and calling patterns.

## III. SYSTEM DESIGN

Our system consists of three main components – as first step it generates the persona or profile of the mobile phone user. The process of this component is illustrated in Figure 1 and described in section III. A. The second system component generates a log correlating two mobile user logs and determining if they were connecting via encrypted messaging services. Figure 2 illustrates this process that is detailed in section III B. The third component focuses on WhatsApp [5] application usage by the mobile user and generates daily and weekly trend graphs from the IPDR logs. We have developed this system using Python 3.7 programming language.

The dataset available to us included the information listed in Table 1. We used all the fields for our analysis. Fields DESTPORT and MSISDN were the most important fields for our analysis along with the instance date and time fields.

### A. Persona Generation

The first component of our system generates the persona or profile of the mobile user. This is derived from the GPRS metadata files. Figure 1 illustrates the various steps of the persona generation process. We also detail these steps below:

**Step 1**: The program is formatted to accept CSV files, so all input files must be in CSV format. The first part of the program opens the file using a with loop.

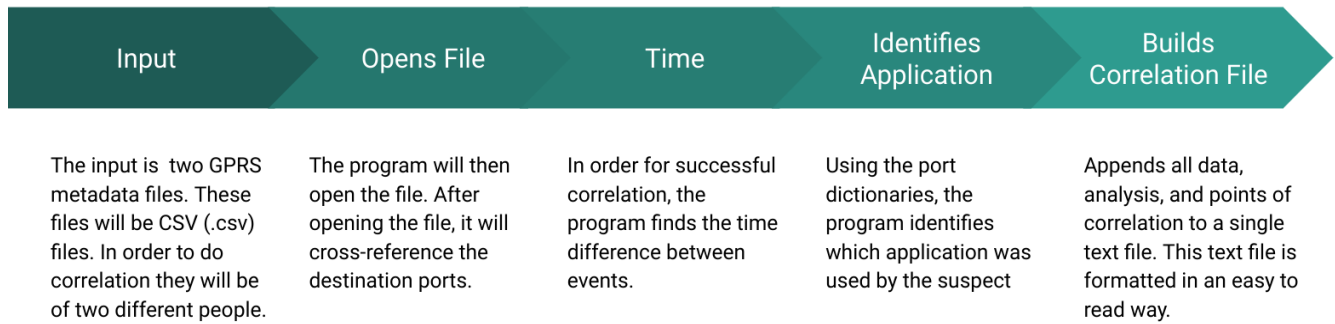

Figure 2: Process to build Correlation File

TABLE I. CDR METADATA FIELDS

| Metadata field | Description |
|---|---|
| PRIVATEIP | IP Address of the Mobile Device |
| PRIVATEPORT | Port of the Mobile Device |
| PUBLICIP | Public IP Address presented by the Mobile Device |
| PUBLICPORT | Public Port presented by the Mobile Device |
| DESTIP | IP Address of the record Destination |
| DESTPORT | Port of the record destination |
| MSISDN | Mobile Station International Subscriber Directory Number - number used to identify a mobile phone number internationally |
| IMSI | International Mobile Subscriber Identity number can identify the user of a mobile network |
| START_DATE, START_TIME | Record Start date and Time |
| END_DATE, END_TIME | Record End Date and Time |
| IMEI | International Mobile Equipment Identity Number that can uniquely identify the mobile phone |
| CELL_ID | ID of the Cell Tower |
| UPLINK_VOLUME | Amount of data Upload |
| DOWNLINK_VOLUME | Amount of data Downloaded |
| TOTAL_VOLUME | Total volume of the Data |
| I_RATTYPE | Identifies whether it is 2G Data or 3G Data |

**Step 2**: After opening the file, program cross references the destination ports. A vital part of this project is the destination port. The destination port is needed to figure out what exactly the suspect is doing on the Wi-Fi. One major obstacle that presented itself was linking destination ports to a destination.

**Step 2A**: In the traditional method, the law enforcement analyst manually extracted all possible destination ports from the data. This was a time consuming process as over nine thousand points of data were analyzed, and destination ports were repeated many times. After this tedious process, one had to link each unique port to a destination. This proved to be the most challenging part, as there is no central database in the public domain with all of the ports and their associations. The process required us to manually browse through multiple websites, some which were helpful and some which were not. Sites that frequently came up were question and answer websites, such as Quora, and Yahoo Answers. For the different applications, we had to manually identify where each port lead to.

**Step 2B**: Using the identified ports, we created lists of the major applications that were frequently accessed. These included WhatsApp, iTunes/QuickTime, Microsoft/Games, and Web access via HTTP and HTTPS. Table II lists all the ports and their corresponding applications that were identified by us.

**Step 3**: The next part identifies the precise destination. Our program runs a reverse DNS search to find the site that was accessed. This is accomplished using the destination IP. If the search is successful, the system returns the web address accessed by the person. Otherwise, the destination IP is returned.

**Step 4**: The next step was to link each port to a destination. Earlier in the process, the ports had already been cross referenced, this step linked each port which existed in the program's dictionary to a destination (see Table II).

**Step 5**: The last step outputs all of the information to a text file in a format that is readable by the user. This output file is used to generate the User profile graph from the data processed. We illustrate such a profile chart in section IV (see Figure 4).

*B. Correlating Callers*

The second component of our system determines the correlation between two mobile users by reviewing their CDR logs and determining the time when the two were potentially accessing the same messaging application. Figure 2 illustrates the process of correlating two mobile callers.

The steps followed by the system to identify the connected mobile users are detailed below:

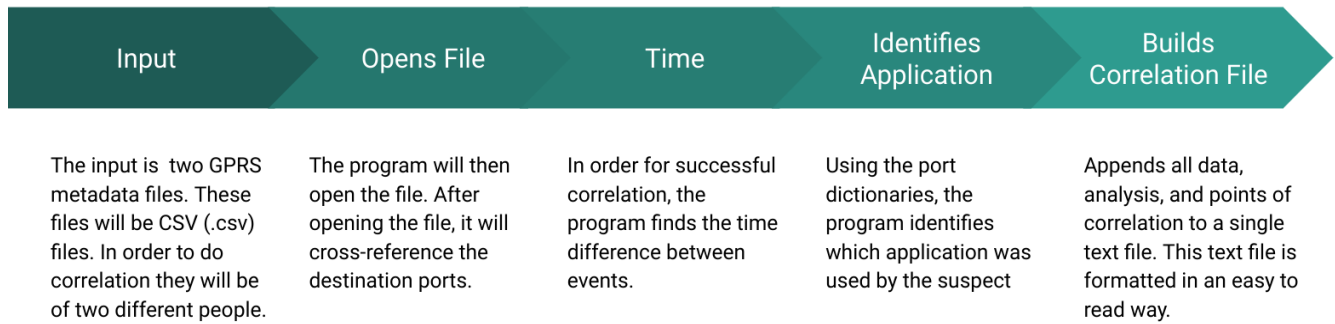

| Input | Opens File | Time | Identifies Application | Builds Correlation File |
|---|---|---|---|---|
| The input is two GPRS metadata files. These files will be CSV (.csv) files. In order to do correlation they will be of two different people. | The program will then open the file. After opening the file, it will cross-reference the destination ports. | In order for successful correlation, the program finds the time difference between events. | Using the port dictionaries, the program identifies which application was used by the suspect | Appends all data, analysis, and points of correlation to a single text file. This text file is formatted in an easy to read way. |

Figure 2: Process to build Correlation File

**Step 1:** To find connections between any two mobile users, we begin with the GPRS metadata files of the users. The files are converted into CSV format and then inputted into the system program.

**Step 2:** The next step is to establish if two events in the GPRS file have the same destination port. All the instances in the two files with the same destination port are identified.

TABLE II. PORT AND APPLICATION LIST

| Destination Port | Description / Protocol | Destination Application | Application Vendor |
|---|---|---|---|
| 5223 | | WhatsApp | Facebook Inc |
| 5228 | | WhatsApp | Facebook Inc |
| 4244 | | WhatsApp | Facebook Inc |
| 5222 | | WhatsApp | Facebook Inc |
| 5242 | | WhatsApp | Facebook Inc |
| 443 | TCP | Skype | Microsoft Inc |
| 3478-3481 | UDP | Skype | Microsoft Inc |
| 49152-65535 | UDP+TCP | Skype | Microsoft Inc |
| 80, 8080 | | Web Connection | |
| 443 | Secure Sockets Layer | Web Connection | |
| 8081 | McAfee E-Policy Orchestrator Black Ice Cap | Web Connection | |
| 993, 143 | | IMAP | |
| 8024, 8027, 8013, 8017, 8003, 7275, 8025, 8009 | Apple iTunes music | iTunes | Apple Inc. |
| 58128, 51637, 61076 | Apple Inc.'s storage area network (SAN) or clustered file system for macOS. | Xsan | Apple Inc. |
| 40020, 40017, 40023, 40019, 40001, 40004, 40034, 40031, 40029, 40005, 40026, 40008, 40032 | | Microsoft Various Online Games | Microsoft Inc. |

**Step 3:** The next part involves determining that the time difference is within the specified threshold. We calculate the difference between the start times of events in the two files that have the same destination port. If the time difference is within the specified threshold, then the program will mark the two as a connection. The threshold time range can be changed at the analyst request.

**Step 4:** The next part is to find out on what application the two people were on at the same time. This is done by linking the destination port to a destination, this is done by referencing the dictionary of ports that was established earlier in table II.

**Step 5:** Once these details are gathered, the system appends them to a file. Figure 3 illustrates this output file generated by comparing some CDR logs of two mobile callers. We provide details about how often connections were made and what fractions of connections were in common. The analyst can provide a threshold above which this degree of common connectivity can be assumed to be a connection. There are clear tradeoffs between false positives and false negatives here, and the choice is a function of the tradecraft of the analyst or the policies of the organization.

*C. Caller's WhatsApp Usage Trends*

The third component of our system generates the trends of WhatsApp usage by a mobile user by analyzing their IPDR logs. This component looks at one or more files to generate usage trends only for WhatsApp application. With one input file of IPDR data, the system creates two output files of results. One file, a text file, lists all the individual connections. The second, a csv file, has a break down for the day based on three-hour intervals (i.e. 12am – 3 am). When a directory is inputted into this component, the only result is a csv file with a breakdown of usage trends within three-hour intervals.

Like the other components of the system, in this component too an individual record is identified as a WhatsApp connection using the destination port. If the destination port in a record is a valid WhatsApp port number (see table II), then it is analyzed further. To provide trends for a three-hour interval, the start time of the record is compared with the time intervals and placed within the correct one. The time intervals are logged in their own CSV file which can be further analyzed if

```
Found the following numbers that were using the same application within 3 minutes of each other
Application  Port  Number1       Date        Start Time  End Time   Number2       Date        Start Time  End Time
WhatsApp     5223  919871808000  28/08/2014  19:29:04    19:32:58   985279543584  28/08/2014  19:29:04    19:32:58
WhatsApp     5223  919871808000  28/08/2014  18:29:47    0:02:40    985279543584  28/08/2014  18:29:47    0:14:40
WhatsApp     5223  919871808000  28/08/2014  18:29:47    0:14:40    985279543584  28/08/2014  18:29:47    0:14:40
WhatsApp     5223  919871808000  28/08/2014  18:29:47    0:26:40    985279543584  28/08/2014  18:29:47    0:14:40
WhatsApp     5223  919871808000  28/08/2014  18:29:47    0:38:40    985279543584  28/08/2014  18:29:47    0:14:40
WhatsApp     5223  919871808000  28/08/2014  18:29:47    4:16:10    985279543584  28/08/2014  18:29:47    0:14:40
WhatsApp     5223  919871808000  28/08/2014  18:30:32    20:31:40   985279543584  28/08/2014  18:29:47    0:14:40
HTTPS        443   919871808000  28/08/2014  18:29:16    6:25:09    985279543584  28/08/2014  18:29:16    6:25:09
HTTPS        443   919871808000  28/08/2014  18:29:47    0:02:40    985279543584  28/08/2014  18:29:16    6:25:09
HTTPS        443   919871808000  28/08/2014  18:29:47    0:14:40    985279543584  28/08/2014  18:29:16    6:25:09
HTTPS        443   919871808000  28/08/2014  18:29:47    0:26:40    985279543584  28/08/2014  18:29:16    6:25:09
HTTPS        443   919871808000  28/08/2014  18:29:47    0:38:40    985279543584  28/08/2014  18:29:16    6:25:09
HTTPS        443   919871808000  28/08/2014  18:29:47    4:16:10    985279543584  28/08/2014  18:29:16    6:25:09
HTTPS        443   919871808000  28/08/2014  18:29:47    0:02:40    985279543584  28/08/2014  18:29:16    6:25:09
HTTPS        443   919871808000  28/08/2014  18:29:47    0:14:40    985279543584  28/08/2014  18:29:16    6:25:09
HTTPS        443   919871808000  28/08/2014  18:29:47    0:26:40    985279543584  28/08/2014  18:29:16    6:25:09
HTTPS        443   919871808000  28/08/2014  18:29:47    0:38:40    985279543584  28/08/2014  18:29:16    6:25:09
HTTPS        443   919871808000  28/08/2014  18:29:47    4:16:10    985279543584  28/08/2014  18:29:16    6:25:09
There were 18 instances of overlap in activity between the two numbers.
The two suspects were on WhatsApp together 7 times. This is 0.3888888888888889% of the total connections.
The two suspects were on a secure web connection together 11 times. This is 0.6111111111111112% of the total connections.
Total number of calls were: 110
Execution time was: 0.22302878697713216 minutes
```

Figure 3: Correlation Result file

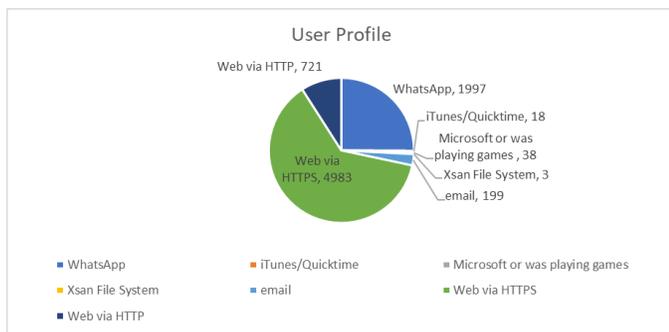

Figure 4: User Profile generated by the system

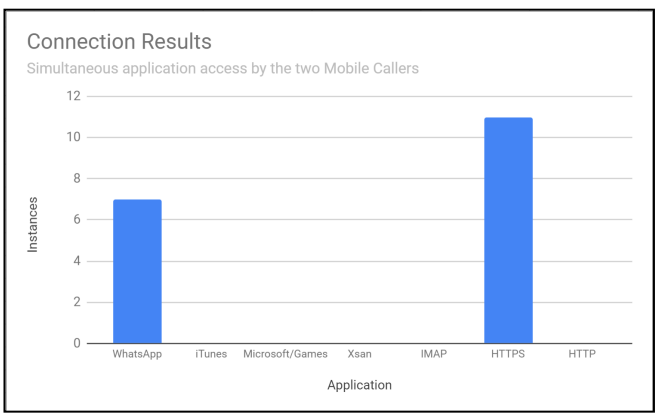

Figure 5: User Connections results generated by the system

needed (see figures 8 and 9). In the appropriate case, the rest of the details are logged in a separate text file (see figure 7).

## IV. SYSTEM EVALUATION

For evaluating our system design, we used the anonymized dataset provided by Delhi Police. This data consisted of CDR and IPDR logs of potential criminals being tracked by the law enforcement agencies.

The results of the persona generation component of our system is illustrated in Table III and Figure 4. It gives an analyst a quick sense of what the suspect is spending time on, what their interests might be etc.

TABLE III.  PERSONA GENERATED FOR MOBILE USER

| Application | Frequency | Usage percent |
|---|---|---|
| WhatsApp | 1997 | 25.09 |
| iTunes/QuickTime | 18 | 0.22 |
| Microsoft or was playing games | 38 | 0.47 |
| Xsan File System | 3 | 0.03 |
| Email | 199 | 2.5 |
| Web via HTTPS | 4983 | 62.6 |
| Web via HTTP | 721 | 9.05 |

The results of the User correlation generation component of our system is illustrated in figure 3 and figure 5. The figure shows how much time correlation takes in a worst-case situation, in this case, with all destination ports matching. Even with 200 data points, it takes about a second. Using the chart editor on Google sheets, we found the equation for the trendline of the data in Figure 6. It is a linear equation. However, since we've naively implemented correlation, we expected the big Oh scaling to be $n^2$.

To validate the accuracy of our work, we manually correlated the data, and then checked those results against the results generated by the system. This was done under multiple different scenarios. The base case was anonymized data from a real suspect's GPRS dump. This was used to create another synthetic GPRS dump which had no common usage with the real suspect dump. This created the first test, and as expected, the system found no overlap. We then created a series of test cases with varying degrees of constructed "overlap". The overlap was also manually verified. This was then compared against the output of our program, and it was found to confirm in each case figure 9 for WhatsApp trends for a week by Time of the day. These graph were found to be very relevant by the law enforcement agencies to determine the time when a suspect is active on the messaging application and help build his profile.

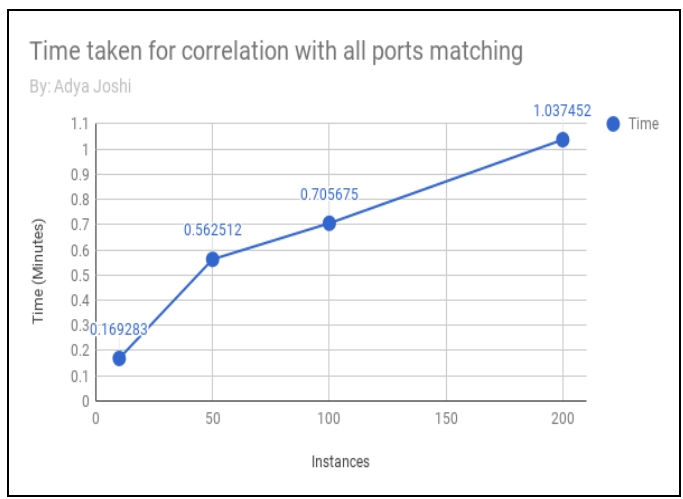

Figure 6: Performance Trends of the correlation

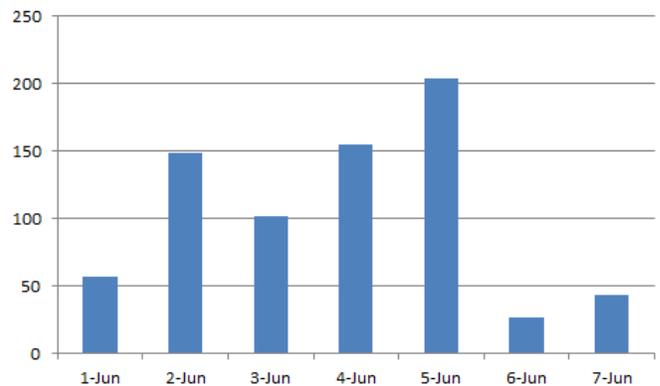

Figure 8: WhatsApp connections trend for a week based on day

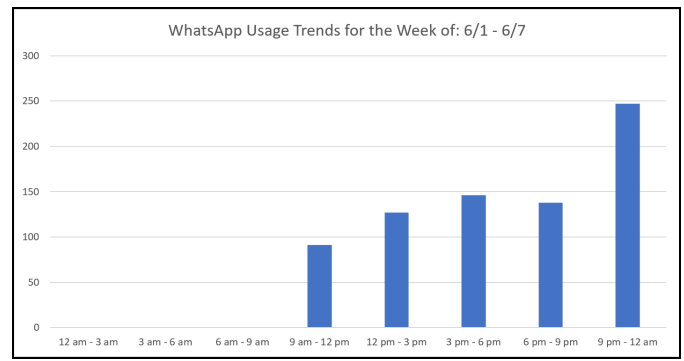

Figure 9: WhatsApp Usage for a week based on Time of Day

![Figure 7 text]

Figure 7: WhatsApp connections identified from the IPDR log

The results of the WhatsApp trends component of our system is illustrated in figures 7,8 and 9. The system output identifies that a connection is being made to WhatsApp based on the port number . (Illustrated in figure 7). Figure 8 illustrates the graphical output for WhatsApp usage for a week by day and

To the best of our knowledge, there are no similar programs that are available publicly. Such programs might exist internally with law enforcement or intelligence communities, but they are not discussed in the public domain.

V. CONCLUSION AND FUTURE WORK

In this paper, we have described the system we developed to generate a profile of a mobile user based on their GPRS CDR metadata. We also developed a technique to determine potential connection between two mobile users based on simultaneous user access of an encrypted messaging service or other similar application ports. This is far less time-consuming than the manual methods in use currently. This system is very useful to the law enforcement agencies who are faced with the challenge of monitoring messaging applications that have end-to-end encryption. Note that this is a proof of concept that demonstrates the power of our approach. Due to legalities associated with data access, we can't directly test our system on multiple real data sets. However, our code has recently been turned over to Delhi Police, who plan to use it in their work.

As part of our on-going work, we are also developing a user-friendly graphical interface for our system. In the future, we plan to enhance this system to allow analysis of other data sets such as Internet Protocol Data Records, IPDRs. We also plan to make some features such as the reverse DNS search optional in an effort to shorten the time taken by the user persona builder.

ACKNOWLEDGEMENT

This work was done while the first author visited Delhi and used the facilities of IIT Delhi and Delhi Police. The first author would like to acknowledge help and guidance provided by Mr. Pradip Kushwaha DANIPS (DCP Special/Technology Cell), Mr. Deepak Vats and Ms. Chanda Sahijwani of Delhi Police, and Mr. Miftah Siddiqui of IIT Delhi.